\documentclass{ceab}   

\usepackage{epsfig}     
\usepackage{graphicx}   

\usepackage{ceabbib}     
\usepackage[T1]{fontenc}

\newcommand{\oc}{(\textrm{O-C})}
\newcommand\hvezda{BS\,Vul}
\def\aut{Z.\,Mikul\'a\v{s}ek, M.\,Zejda, L.\,Zhu, S.-B.\,Qian, J.\,Li\v{s}ka,
S.\,N.\,de Villiers:}
\def\tit{EB visual minima timings: To use or not to use?}

\begin{document}

\title{Visual minima timings of eclipsing binaries: To~use or not to use?}

\author{Z. MIKUL\'A\v SEK$^{1,2}$, M. ZEJDA$^1$, L. ZHU$^3$, S.-B. QIAN$^3$,
J. LI\v SKA$^{1}$,\\ and S.~N. de~VILLIERS$^{4}$
\vspace{2mm}\\
\it $^1$Department of Theoretical Physics and Astrophysics,
Masaryk University,\\
\it Kotl\'a\v rsk\'a 2, CZ-611~37 Brno, Czech Republic, mikulas@physics.muni.cz \\
\it $^2$Observatory and Planetarium of Johann Palisa, V\v SB -\\
\it Technical University, Ostrava, Czech Republic\\
\it$^3$National Astronomical Observatories/Yunnan Observatory, \\
\it Chinese Academy of Sciences, Kunming, China\\
\it $^4$Priv. Observatory, 61 Dick Burton Road, Plumstead, Cape Town,
South Africa}

\maketitle

\begin{abstract}
Plenty of mid-eclipse timings of short-periodic eclipsing binaries
derived from series of visual observations appear to be an acceptable
source of information for orbital period analyses, namely if they were
done in time-intervals not covered by other types of observations.
However, our thorough period analysis of the nearly contact eclipsing
binary BS\,Vulpeculae proves that visually determined timings done in
1979-2003 were undoubtedly biased to accommodate the existing linear
ephemeris. The heavily subjective character of visual observations
disqualifies them as a source of true phase information apt for fine
eclipsing binary period analyses. Consequently we warn against the use
of visual timings without a preceding careful verification.
\end{abstract}

\keywords{eclipsing binaries - minima timings - visual observation -
BS Vul}

\section{Introduction}

There are many good reasons for the analysis of orbital periods of
eclipsing binaries hereafter EBs). Many of the mysterious,
astrophysically interesting binaries were revealed thanks to the
analysis of their odd orbital period variations. Fine period analysis
provide additional information about variable stars physics. By means
of period analysis, it is possible to reveal and characterise the
existence of further bodies in the EB system (stars, planets) or mass
exchange between interacting binary components. By analysing period
variations caused by apsidal motion we can deduce the internal
structure of the components and test the theories of gravity. From the
systematic decrease of the period of close EBs, it is possible to
reason on the angular momentum loss through stellar winds or
gravitational waves, etc.

These period variations are always rather delicate. The credibility
of~conclusions based on EB period analysis depends strongly on the
applied methods and reliability of the data used.

\section{Data for period analyses of eclipsing binaries}

There are two types of EB data, used together with phase information,
that are standardly used for orbital period analysis -- original time
series of EB measurements (A) and mid-eclipse timings (B).

\noindent \textbf{A}. Various EB measurement time series types in the
usual order of reliability are:
\begin{itemize}
\item{photoelectric or CCD time series of parts of light curves done
in several filters where EB minima are targeted or non-targeted;}
\item{brightness measurements done by photometric surveys non-targeted
to the particular object (ASAS, Hipparcos etc.);}
\item{photoelectric or CCD time series of parts of light curves done
in one filter or integral light where EB minima are targeted as a
rule;}
\item{time series of spectroscopic observations (line strengths,
radial velocities etc.);}
\item{time series of magnitudes derived from archival photographic
plates;}
\item{time series of individual visual estimates of EB brightness
where eclipses are not targeted.}
\end{itemize}

\noindent \textbf{B}. Timings of primary/secondary EB minima are
preprocessed data derived from original time series of type A done as
a rule during central parts of eclipses. The whole time series is then
degenerated into only one result -- time of the mid eclipse -- derived
standardly by an obsolete \citet{kwee} method. This method yields a
usable estimate of minima times if descending and ascending branches
are of the same length. Nevertheless, the uncertainties of minima
timings were generally underestimated and consequently unacceptable.

Here follows eclipsing binary minimum light timing sources in order of
their reliability:
\begin{itemize}
\item{EB timings derived from photoelectric or CCD time series (a
voluntary degradation from category A to B);}
\item{timings when the EB was definitely fainter than in phases
outside eclipses - an objective source of information, but of low
quality;}
\item{plenty of amateur visual observations of EB minima - a subjective
source of information of problematic quality. Previously this was a
most popular and respected amateur activity with a clear scientific
output. The subjective source of information with problematic
quality.}
\end{itemize}

\noindent The only advantage of mid-eclipse timings (B) is the
homogeneity of data from different observers and it is the main reason
why the majority of period analyses were done using these types of
data. However, a lot of such analyses are erroneous because of lowered
reliability of entering data and the fact that times of minima almost
do not use the fact that the light curves are periodical functions.
Our experience is that the results of direct period analysis, based on
time series of EB measurements (A), are at least twice as accurate.

\section{Subjective nature of visual estimates of EB magnitudes}

From our long, in-depth experience with visual observations of EBs it
became clear that these data should be treated with caution since the
data suffer from many bad attributes that effectively limit their
suitability for fine period analysis purposes. We can enumerate only
the most serious ones:
\begin{itemize}
\item Due to the poor quality of visual observations, the accuracy of
the determined time of minimum light is much worse than for other
types of photometric observations. The quality of visual timings are
further deteriorated by the fact that some determinations were based
on only a few individual observations done only in the immediate
vicinity of the predicted time of minima (the majority of visual
observation runs begin one hour before the expected minimum time and
end one hour after it). The extreme shortness of observation runs are
typical for some observers and one should be careful when using such
unreliable minima timings \citep[see the number of estimates used for
minima timing determinations in the representative list of such data
in][]{zhu}.
\item For time series of visual estimates, minimum light times were
often determined by obscure, irreproducible methods. In the
worst case, the original estimates were not published and they are now
inaccessible.
\item Almost all standard methods used to estimate mid-eclipse
timings are based on the concepts of the least square method. However,
this method strictly requires successive estimates to be independent.
This requirement is definitely not kept in time series of visual
observations. A`visual observer' is not an instrument but a subject
who knows what light curves of observed EBs should look like. He
remembers his last estimates and subconsciously strives to create the
most plausible light curve. `Observed' visual light curves are thus
subjectively smoothed to the ideal of theoretical or photoelectric
light curves with clearly defined descending and ascending branches.
This smoothing i.a. prevents us from estimating the real uncertainty
of the timings' determination from the time series of only one
observing session. The uncertainties formally calculated and published
from the smoothed light curve data are many times smaller than they
should be.
\item The most severe flaw in the majority of the visual
observations is that observers obviously knew the predicted time of
minimum light to an accuracy of minutes. If the ephemeris for the time
of minimum light was incorrect, most observers were influenced into
confirming the predicted, incorrect time of minimum light. This
subjective effect, which is able to completely invalidate
observations, equally afflicts both beginners and `experienced' visual
observers.
\end{itemize}
All these statements can be exemplary documented for the case of
eclipsing binary BS Vulpeculae.

\begin{figure}
\begin{center}
   \includegraphics[scale=0.7,angle=-0]{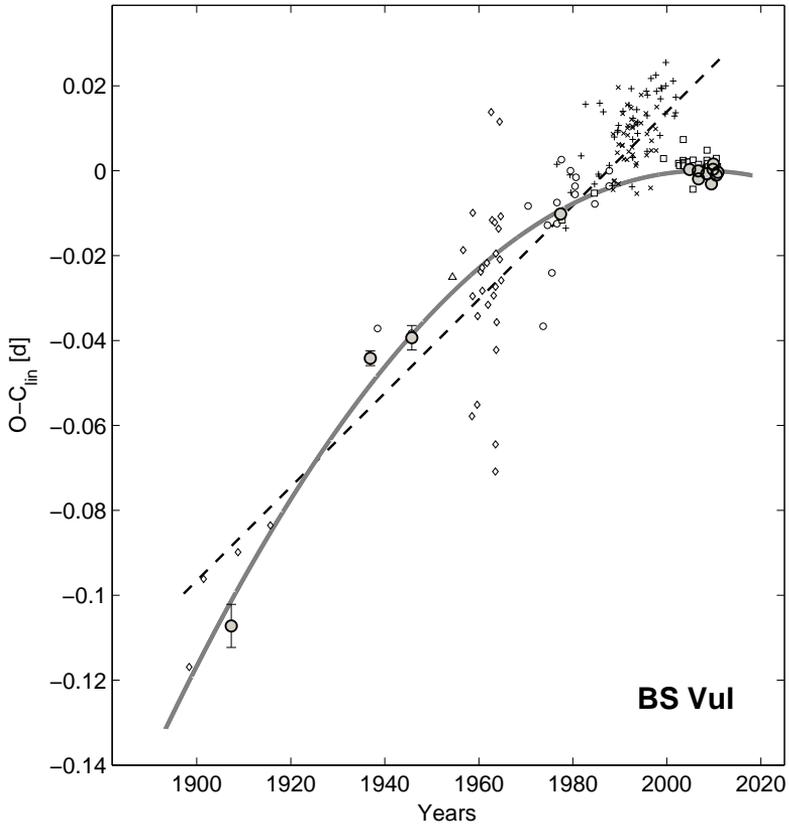}
\end{center}
\caption{\oc\ diagram with respect to the linear ephemeris of primary
minima times: JD$_\mathrm{I}=2454387.8692+0.47597003\times E$, where
$E$ is an integer (epoch). Large circles with error bars correspond to
`virtual times of eclipses' \citep[][]{mikbezoc}. The small symbols
`+' show published minima timings of visual observations of G.
Samolyk, `$\times$' denote timings of H. Peter, and small circles
correspond to minima obtained by other visual observers. Small
diamonds are timings of photographic plates on which the star is close
to minimum and small squares are photoelectric or CCD minima timings.
The dashed line corresponds to the linear ephemeris of \citet{bernard}
JD$_\mathrm{I}=2443271.578+0.47597147\times E$, used by visual
observers. The bold grey line corresponds to the found quadratic
ephemeris.}\label{fig1}
\end{figure}

\section{BS Vulpeculae}
\hvezda\ is a relatively bright, but neglected nearly contact EB with
very short orbital period $P \sim 0.476$\,d. \citet{zhu} recently
analysed the period by the `direct' period analysis method
\citep[briefly described in][]{mikbezoc} using 8177 individual
brightness observations of \hvezda\ done during the time-interval
1898-2010. They found that the period is shortening. The result proves
that \hvezda\ evolves toward contact phase.

Later we improved the method used in \citet{zhu} so that it is now
possible to combine all the above mentioned data with 46 mid-eclipse
timings derived from time-series obtained by objective photometric
measurements (photoelectric photometry, CCD photometry or acquiring of
photographic snaps series) in a time vicinity of the predicted minima
listed in \citet{zhu}. We confirmed the shortening of the orbital
period and improved its value to $\dot{P}=-2.11(5)$ ms/yr (see also
\oc\ diagram in Fig.\,\ref{fig1}).

\subsection{\hvezda\ visual minima timings}

The sufficiently deep primary eclipse (in visual 0.7 mag) and short
duration (only 3.5\,h) predestined \hvezda\ to be an attractive target
for visual observers. We found in literature 104 primary mid-eclipse
timings derived from visual observations in the time-interval
1970--2003 \citep[see the list in][]{zhu}. The large number of visual
timings enables a relatively detailed discussion of their properties.

\begin{figure}
\begin{center}\includegraphics[width=0.9\textwidth]{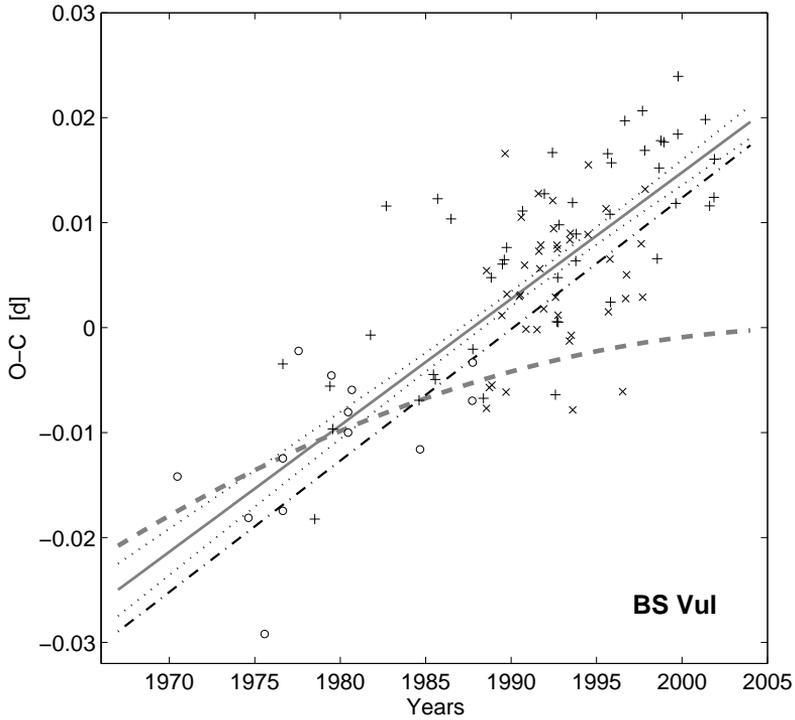}
\end{center}
\caption{\oc\ diagram of visually determined primary eclipse timings
with respect to the linear ephemeris
$\textit{JD}_\mathrm{I}=2454387.8692+0.47597003\times E$, where $E$ is
an integer (epoch). The symbols `+' show G. Samolyks' timings,
`$\times$' denote H. Peters' timings, and circles correspond to
timings obtained by other visual observers. The dashed-dot line
corresponds to the linear ephemeris of \citet{bernard}
$\textit{JD}_\mathrm{I}=2443271.578+0.47597147\times E$, used by
visual observers. The dashed curve is the real quadratic course. The
full line is the linear fit of visually determined timings; dotted
lines are the 1-$\sigma$ uncertainty of this fit.}\label{fig2}
\end{figure}

Our analysis of all 104 visual timings shows their excellent
correlation with the linear ephemeris of \citet{bernard}
JD$_\mathrm{I}=2443271.578+0.47597147\times E$, that was generally
used by visual observers for their observation schedule
(Fig.\,\ref{fig2}). The linear fit of visual observations is almost
parallel with the line of prediction, while timings of minima
systematically lag behind the minimum forecast by several minutes.
Visual observations have only a very weak correlation with the  real
quadratic ephemeris. The conclusion is obviously valid for almost all
visual observers who targeted the star. It is apparent that
practically all visual observers subjectively adjusted their
observations to more or less confirm the existing ephemeris. The
systematic deflections had grown to 26 minutes by 2003.

The apparent `delay' in visual observers' timings with respect to
prediction agrees well with the heliocentric corrections for the times
of observation. Observers used geocentric time, but the predicted
minimum light time was  heliocentric, which was (in the case of
\hvezda) delayed several minutes with respect of the geocentric time
on the night when observations were done.

The most active visual observer of \hvezda\ G. Samolyk, published also
528 brightness estimates (accessible through the AAVSO webpage)
obtained during his observations of 39 primary eclipses of \hvezda\ in
the time interval 1984--2002.

We analysed all these data in detail and arrived i.a. at the following
conclusions:

\begin{figure}
\begin{center}
   \includegraphics[scale=0.76,angle=-0]{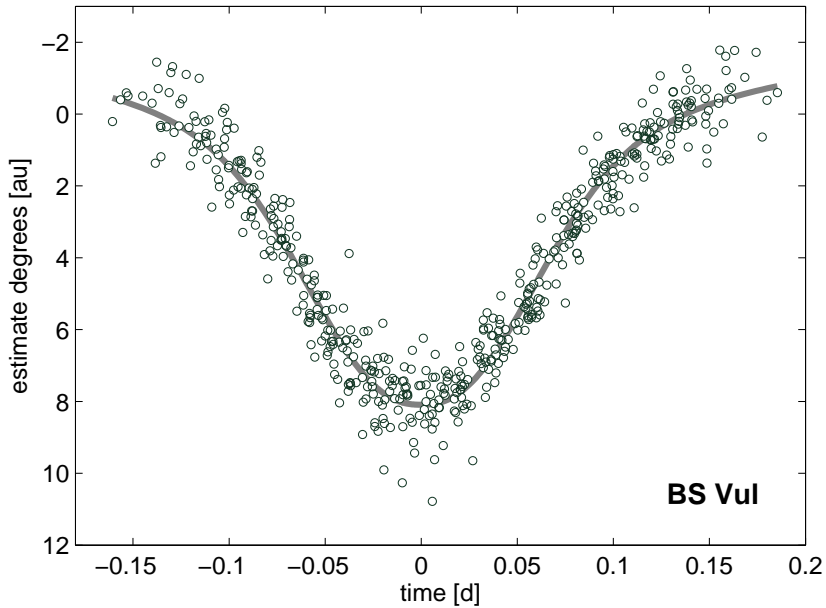}
\end{center}
\caption{Samolyk's mean light curve of the \hvezda\ primary minimum
depicting his brightness estimates in arbitrary units versus the times
of minima determined by him looks very authentically.}\label{fig3}
\end{figure}

\begin{itemize}
\item Observations of \hvezda\ were (due to its extremely short period)
always optimally scheduled, typically 12 estimates per eclipse were
able to describe the complete eclipse quite well. Both descending and
ascending branches of the light curve during eclipses were covered
nicely.
\item The resulting light curves were perfectly symmetric and they
corresponded very well with ideal photometrically acquired curves (see
Fig.\,\ref{fig3}). Samolyks' light curves were very smooth; their
scatter derived from the form of the light curve is only 0.055~mag!
However, the scatter of individual times of eclipses indicates that
the real scatter in estimates should be at least five times larger!
\item The scatter of timings published by Samolyk and their bad
affinity to the real quadratic ephemeris might have been caused by the
application of an inappropriate method for the derivation of light
curve minima from estimate series. Since we know the light curve
shape, we applied our own rigorous method \citep[][]{mikproc} to
determine the times of minima. Although we obtained diverse times of
minima, all of them were clustered around the \citet{bernard}
`geocentric' ephemeris. Consequently, the estimates themselves were
biased to accommodate the forecast.
\end{itemize}
It is evident from our findings that the heavily subjective character
of visual observations of short period EBs disqualifies them as a
source of unbiased information apt for fine EB period analysis. On the
other hand we could use 200 visual estimates of brightness made by S.
Piotrowski in two time intervals 1935--1939 (155 estimates) and
1945--1946 (45 estimates) \citep[data in][]{aca}, because these
estimates have a different character than later AAVSO and BBSAG visual
observations. The distribution of these old visual estimates shows
that they were obtained in `monitoring mode' without the primary aim
to obtain minima timings. Furthermore they passed our careful
analysis.

\section{Conclusion}
Our analysis of mid-eclipse timings of \hvezda\ derived from the time
series of visual observations proved the heavily subjective character
of visual observations of \hvezda. It should be noted that mid-eclipse
timings derived from observations of visual observers associated with
B.R.N.O in pre-email times were not biased by the knowledge of the
expected minima times of EBs. The reason for this is that they used
special forecasts of EB minima distributed from the centre where times
were rounded to the nearest half hour.

We conclude that the heavily subjective character of visual
observations disqualifies visual minima timings as a source of true
phase information necessary for fine eclipsing binaries period
analyses. Consequently we do not advise using these data without
having done a detailed verification of their reliability.

\section*{Acknowledgements}
We wish to thank GA\v{C}R 209/12/0217, MUNI/A/0968/2009, ME10099, and
LH12175. This work was partly supported by the intergovernmental
cooperation project between P.\,R.\,China and the Czech Republic.

\bibliographystyle{ceab}

\section*{References}
\begin{itemize}
\small
\itemsep -2pt
\itemindent -20pt
\bibitem[de Bernardi \& Scaltriti(1979)]{bernard} de Bernardi, C., and
    Scaltriti, F.: 1979, {\it Astronomy and Astrophysics, Suppl. Ser.} {\bf 35}, 63
\bibitem[Kwee \& van Woerden(1956)]{kwee} Kwee, K.~K., and van Woerden,
    H.: 1956, {\it Bulletin of the Astronomical Institutes of the
    Netherlands} {\bf 12}, 327
\bibitem[Mikul\'a\v sek et al.(2012a)]{mikbezoc} Mikul\'a\v sek, Z.,
    Zejda, M., and Jan\'ik, J.: 2012a, in \emph{From Interacting Binaries
    to Exoplanets: Essential Modeling Tools}, IAUS 282,
    Eds. M. Richards \& I. Hubeny, 18--22 July 2011, Tatranska
    Lomnica, Slovak Republic, 391
\bibitem[Mikul\'a\v sek et al.(2012b)]{mikproc} Mikul\'a\v sek, Z.,
    Zejda, M., Qian, S.-B., Zhu, L., 2012b, in \emph{Proceedings of 9-th
    Pacific Rim Conference on Stellar Astrophysics}, 14--20 April 2011,
    Lijiang, China, ASP Conference Series, 111
\bibitem[Szafraniec(1962)]{aca}Szafraniec, R.: 1962, {\it Acta Astronomica Suppl.}
    {\bf 5}
\bibitem[Zhu et al.(2012)]{zhu}Zhu, L.-Y., Zejda, M., Mikul\'a\v sek, Z., Li\v ska, J.,
    Qian, S.-B., de Villiers, S.~N.: 2012, {\it Astronomical Journal}
    {\bf 144}, 37.
\end{itemize}

\end{document}